\documentclass[runningheads]{llncs}
\usepackage{cite}
\usepackage{tabularx}
\usepackage{amsmath,amssymb,amsfonts}
\usepackage{algorithmic}
\usepackage{graphicx}
\usepackage{csquotes}
\usepackage{textcomp}
\usepackage[table]{xcolor}

\usepackage{colortbl}
\usepackage{array,tabularx}
\definecolor{tablerow1}{RGB}{230,230,230}
\definecolor{tablerow2}{RGB}{245,245,245}

\usepackage[most]{tcolorbox}
\begin{document}
\title{Generative AI in Undergraduate Information Technology Education - Insights from nine courses}
%
%
\author{Anh Nguyen Duc\inst{1}, Tor Lønnestad\inst{1}, Ingrid Sundbø\inst{1}, 
Marius Rohde Johannessen\inst{1}, Veralia Gabriela\inst{1}, Salah Uddin Ahmed\inst{1} 
and Rania El-Gazzar\inst{2}}
\authorrunning{Paper to be presented at NOKOBIT 2023}
%
\institute{University of South Eastern Norway, Bø i Telemark, Norway \\
\email{\{angu, tor.lonnestad, ingrid.sundbo, marius.johannessen, veralia.g.sanchez,salah.ahmed\}@usn.no}
\and
University of Agder, Kristiansand, Norway \\
\email{rania.f.el-gazzar@uia.no}}
\maketitle              
\begin{abstract}
The increasing use of digital teaching and emerging technologies, particularly AI-based tools, such as ChatGPT, is presenting an inevitable and significant impact on higher education. The capability of processing and generating text could bring change to several areas, such as learning assessments or learning experiences. Besides the negative impact, i.e exam cheating, we also see a positive side that ChatGPT can bring to education. This research article aims to contribute to the current debate on ChatGPT by systematic reflection and experience reported from nine bachelor IT courses at a Norwegian university. We conducted inductive empirical research with reflective notes and focused groups of lecturers from nine different IT courses. The findings were thematically organized with numerous use cases in teaching IT subjects. Our discussion highlights the disruptive implications of AI assistant usage in higher education and emphasizes the need for educators to shape this transformation 

\keywords{large language models \and prompt patterns \and  prompt engineering \and software engineering \and ChatGPT}
\end{abstract}

\section{Introduction}
In the era of modern education, technology has ushered in a transformative wave that has redefined the way we teach and learn. Among these innovations, i.e. gamification, virtual reality and distant learning, Generative Artificial Intelligence (AI) has emerged as a promising tool that transcends the boundaries of traditional pedagogy. Generative Artificial intelligence (GenAI) generates different types of content, such as pictures, text, audio, and 3D models, based on the input they receive. Famous examples of GenAI include ChatGPT and Google's Bard  (Google). ChatGPT, a large language model based on GPT-3, is an AI chatbot designed to understand and respond to a wide variety of questions and topics, which has the potential to revolutionize our interaction with technology \cite{haque2022}. While some view ChatGPT as a threats to traditional systems that should be prohibited \cite{dwivedi2023,hamdan2023,lund2023a,lund2023b,ventayen2023}, we cannot ignore the potential of efficiency and productivity the tool could bring to our society. For instance, it could be used for voice user interfaces to overcome issues with response behavior or response quality \cite{klein2021}. AI-based tools can support the creation of new ideas or help in terms of automating tasks. 

In the education domain, the emergence of AI has an already direct impact on higher education, including unknown handling by students, more time-consuming assessments, and the need to evaluate exams differently due to established control structures such as plagiarism checkers and AI detection tools not being able to distinguish between AI-generated and human-generated text. As a result, we will have a change in the way we conduct and evaluate exams. These examples will not be the only changes. Equally important is a change at the program level toward how student-centered style, activities, and approaches can be supported by this tool.

We believe this will be a revolutionary shift in IT education. Thus, it is important to have discussion and guidance on how to deal with such emerging technologies to actively shape this transformation. We contribute to the discussion about ChatGPT in education by systematically examining the adoption of ChatGPT in several courses in Information Technology bachelor programs at the University of South Eastern Norway (USN). Our discussion is associated with different types of courses, from theory courses, programming courses and project-based courses. Besides reflective notes from course lecturers, we also gathered experiences from courses where ChatGPT has been used for teaching, supervising and assessment. We present these reflections under four categories: strengths, weaknesses, threats, and opportunities. We enriched the discussion with evidence from related literature.

The paper is organized as follows: Section 2 is background and related work. Section 3 presents our research method. Section 4 briefly describes the setting of our courses. Section 5 presents our findings. Section 6 discusses and concludes the paper.

\section{Background}
\subsection{Generative AI and ChatGPT}
Large Language Models (LLMs) are artificial intelligence systems trained to understand and process natural language. They are a type of machine learning model that uses very deep artificial neural networks and can process vast amounts of language data \cite{macneil2022}. An LLM can be trained on a very large dataset comprising millions of sentences and words. At the moment, the most popular LLMs include BERT, T5 and GPT-3 \cite{lund2023a,lund2023b}.
Generative modeling is an artificial intelligence (AI) technique that generates synthetic artifacts by analyzing training examples, learning their patterns and distribution \cite{jovanovic2022}. Generative Pre-trained Transformer (GPT) models generate text in different languages and can create human-sounding words, sentences, and paragraphs on almost any topic and writing style \cite{karpathy2015}. One recent example is ChatGPT, a large language model trained by OpenAI \cite{brown2020}. It is designed to generate human-like text based on a given prompt or context. It can be used for various natural language processing tasks, such as text completion, conversation generation, and language translation \cite{radford2018}. When given a prompt or context, the model processes the input and generates a response. The response is generated one word at a time, with the model predicting the next word based on the input and the words it has generated so far. After the training, the model can be fine-tuned on a specific task, such as question answering or dialogue generation, by providing it with task-specific examples and fine-tuning the model on this data. It can also generate text in multiple languages by fine-tuning the model on the multilingual dataset or by providing the model with the language code (ChatGPT 2023).
This research discusses ChatGPT 3 due to its current availability in Norway, which has made it a focal point of interest. ChatGPT-3 uses a deep neural network architecture known as Transformer. The model is trained on a massive dataset of text from the internet, which allows it to generate human-like responses to a wide range of input prompts. At a high level, the model works by taking in a sequence of input tokens (usually words or sub-words) and using them to predict the next token in the sequence. For an incomplete sentence, for example: “The book is on the”, these models use training data to produce a probability distribution to determine the most likely next word, e.g., ”table” or ”bookshelf”.
\subsection{Some recent work on GenAI in education}
Much education research has been done on topics like Artificial Intelligence in general, personalized learning and chatbots within the context of higher education. For instance, Wartman and Combs argued that education is evolving in lockstep with changes in the professional sector, demanding the use of Artificial Intelligence (AI) in teaching and learning \cite{wartman2018}. Alam and Mohanty conducted a systematic literature review to summarize ethical considerations, challenges, and potential threats linked to the application of AI in higher education while also exploring potential uses of AI \cite{alam2022}. Their findings were grouped into four categories: intelligent tutoring systems, personalization and adaptive systems, evaluation and assessment, and prediction and profiling. Their research highlights a shortage of critical thinking regarding the challenges and potential threats of using AI in higher education. A chatbot is a computer program that is capable of maintaining a conversation with a user in natural language, understanding their intent, and replying based on preset rules and data \cite{adamopoulou2020}. Okonkwo and colleagues conducted a systematic literature review of 53 articles about chatbot applications in education \cite{okonkwo2021}. Among common use cases, chatbot systems were reported to deliver course content to students via an online platform as a conversational agent capable of providing accurate information to users and getting individualized help.

Since the release of ChatGPT in November 2022, there have already been several publications on AI-assisted education. For example, Rudolph et al. have examined potential use cases in higher education and recommend an engaged approach where educators figure out ways of incorporating AI assistant technology in teaching and examinations \cite{rudolph2023}. Kung has performed experiments using ChatGPT to answer medical exams and suggests the technology has the potential to assist in medical education \cite{kung2023}. Pavlik has done a similar experiment for media and journalism education and found that AI can be a decent co-author. His paper was written with ChatGPT as a “co-author” \cite{pavlik2023}. Baidoo-Anu et al. discuss how ChatGPT can assist in the learning process and also some of the potential pitfalls, such as wrong factual information, incomplete data sets, or bias in data training \cite{baidooanu2023}. Most recently, Nguyen-Duc et al. conducted a comprehensive survey of more than 200 recent works on GenAI in software engineering and education, and presented some future directions for GenAI in software engineering education \cite{nguyen-duc_generative_2023}

\section{Research method}
Lecturers from nine IT courses were invited to participate in the study. Before group meetings, each of them wrote a reflection note about the application of ChatGPT in their courses. They were given a list of questions to guide their reflections (Table 1). The results were a written collection of the participant’s responses. We followed up with focus groups, as it allows researchers to collect a large amount of data from a substantial group of people in a relatively short amount of time, particularly for exploring how people perceive, feel about, or view a certain domain \cite{wilson2012}. Two focus groups were conducted online during March-April 2023. The focus groups were conducted as discussion sessions, each lasting 60 to 90 minutes \cite{wilson2012}. The implementation was conducted by involving nine lecturers, including the moderator in the research group. The first session focused on generating and synthesizing the themes regarding the research question. The second session adopted the interpretation approach in the context of self-understanding, where they tried to formulate in a condensed form what the participants themselves meant and understood (Spencer et al., 2003)
 
To guide the analysis step, we used the SWOT framework \cite{leigh_swot_2009}. SWOT presents Strengths, Weaknesses, Opportunities, and Threats related to a plan, situation, business competition, or project planning. It is also called situational assessment or situational analysis. Strengths and weaknesses are internal to the organization and can be directly managed by it, while the opportunities and threats are external and can only anticipate and react to them. The framework is also commonly used in academics for analyzing teaching and learning situations.

\begin{table}
\centering
\caption{Reflective questions used in SWOT regarding the adoption of ChatGPT}
\label{tab:my_table}
\begin{tabularx}{\textwidth}{l X}
\hline
\textbf{Section} & \textbf{Questions} \\
\hline
\textbf{Strengths} & 1. What are the advantages of adopting ChatGPT in your course? \newline
2. What unique features does ChatGPT offer that can enhance the teaching and learning experience in your course? \newline
3. What are the potential benefits for students, in your course? \\
\hline
\textbf{Weaknesses} & 4. What are the disadvantages of adopting ChatGPT in your course? \newline
5. What are the potential limitations of ChatGPT in terms of domain knowledge in your course? \newline
6. What are the potential negative impacts on student-teacher interaction in your course? \newline
7. What are other potential technical challenges? \\
\hline
\textbf{Threats} & 8. What are the potential negative impacts on the quality of the course? \newline
9. What are the potential negative perceptions or resistance from students, teachers, or parents towards using ChatGPT in your course? \\
\hline
\textbf{Opportunities} & 10. What are the potential opportunities to customize ChatGPT to meet the specific needs of students and teachers in your course? \newline
11. What are the untapped areas in education where ChatGPT can be applied, such as language teaching or special needs education? \newline
12. How can ChatGPT be integrated into existing pedagogical approaches to create new teaching and learning experiences that were not previously possible? \\
\hline
\end{tabularx}
\end{table}

\section{Study context} 
The Bachelor in IT and Information Systems offered by the University of South-Eastern Norway provides students with a comprehensive education in various areas of information technology. This program is designed to equip students with practical skills in program and system development, web-based solutions, user support, and IT management. With a focus on both theory and practical application, students will gain a deep understanding of the latest technological developments in the field while also developing the problem-solving skills required to excel in a constantly changing industry. By the end of the program, graduates will have the knowledge and expertise to pursue a successful career in various sectors of the IT industry. The bachelor's program is in Norwegian. We categorize the courses into three categories:
\begin{itemize}
    \item Programming courses: Web development and Human-Computer Interaction (WEB1000) and Database (DAT1000). WEB1000 teaches students basics of HTML, CSS, and JavaScript. DAT1000 teaches the student the basics of database system’s concepts, terminology, database management systems, structured query language (SQL) and relational models (ER).
    \item Theory courses: Information System (INF1000), Artifical Intelligence for Business Applications (AI3000), and Digital Strategy (DIG3000). INF1000 provides students with an introduction to the field of Information Systems and current research in the field. AI3000 focuses on the impact of Artificial Intelligence (AI) on organizations and society in general. DIG3000 provides students with an introduction to the strategic use of IT in organizations and how organizations can use technology to support new and existing business processes and develop new business models.
    \item Project-based courses: Practical project work (PRO1000), Business Intelligence and Data Warehouse (BID3000), and System Development (SYS1000). PRO1000 focuses on software project work and project development methodology, with an emphasis on Agile development. In BID3000, students learn about the ETL-process (extraction - transformation - lasting), the data cube, and how to plan the introduction of a BI system. SYS1000 is an introductory course on software development covering project management, processes, models, methods, techniques, and tools.
\end{itemize}

\section{Results} 
The result is organized into four aspects of adopting ChatGPT – strength (Section 5.1), weakness (section 5.2), opportunities (section 5.3) and threat (Section 5.4).
\subsection{Strength}
ChatGPT can be a powerful tool for teaching IT subjects to students for a variety of reasons. ChatGPT can support the formulation of structured theoretical elements, providing examples to demonstrate for cases, concepts and phenomena. Another strength is its ability to provide quick and accurate responses to a wide range of questions, which can be particularly helpful in programming courses where students may encounter abstract and inter-related topics. Moreover, ChatGPT can adapt its responses to the specific needs and learning styles of individual students, providing a personalized learning experience. It is common that some students are fast to learn and some others are slower. A ChatGPT-based virtual assistant can be accessed from anywhere at any time, making it a convenient resource for students who need assistance outside of class hours or who are studying remotely.

As a multi-language chatbot, ChatGPT can remove language barriers for non-native English-speaking students, making IT education more accessible to a wider range of learners. For programming courses, learning materials, including technique description, concept explanation, code comment, etc., originally in English, can be easily offered in a local language.  It can also be helpful in generating ideas and starting points for a lecture, assisting with problem-solving, and even acting as a conversational partner for students who feel lonely in classes. The tool can be programmed to offer encouragement and positive reinforcement, helping to boost their confidence and morale. If a student is struggling with a particular concept, ChatGPT can suggest additional resources or provide examples to help them better understand the topic.
In a project-based course, ChatGPT can be a valuable resource for project management and development. It can assist with finding reading materials for a subject and explain general domain or contexts. For example, if a student is working on a project related to e-commerce, ChatGPT can explain the basics of e-commerce, including different models, challenges, and opportunities. ChatGPT can assist in generating ideas for team and task assignments. For example, if a team is working on a project related to building a mobile app, ChatGPT can suggest typical architectural models, features and functionalities that can be included in the app. If a team practices Agile methodology, ChatGPT can act as a source of information about Agile principles, techniques, best practices and describe how to implement it effectively.
\begin{figure}
    \centering
    \includegraphics[width=0.8\linewidth]{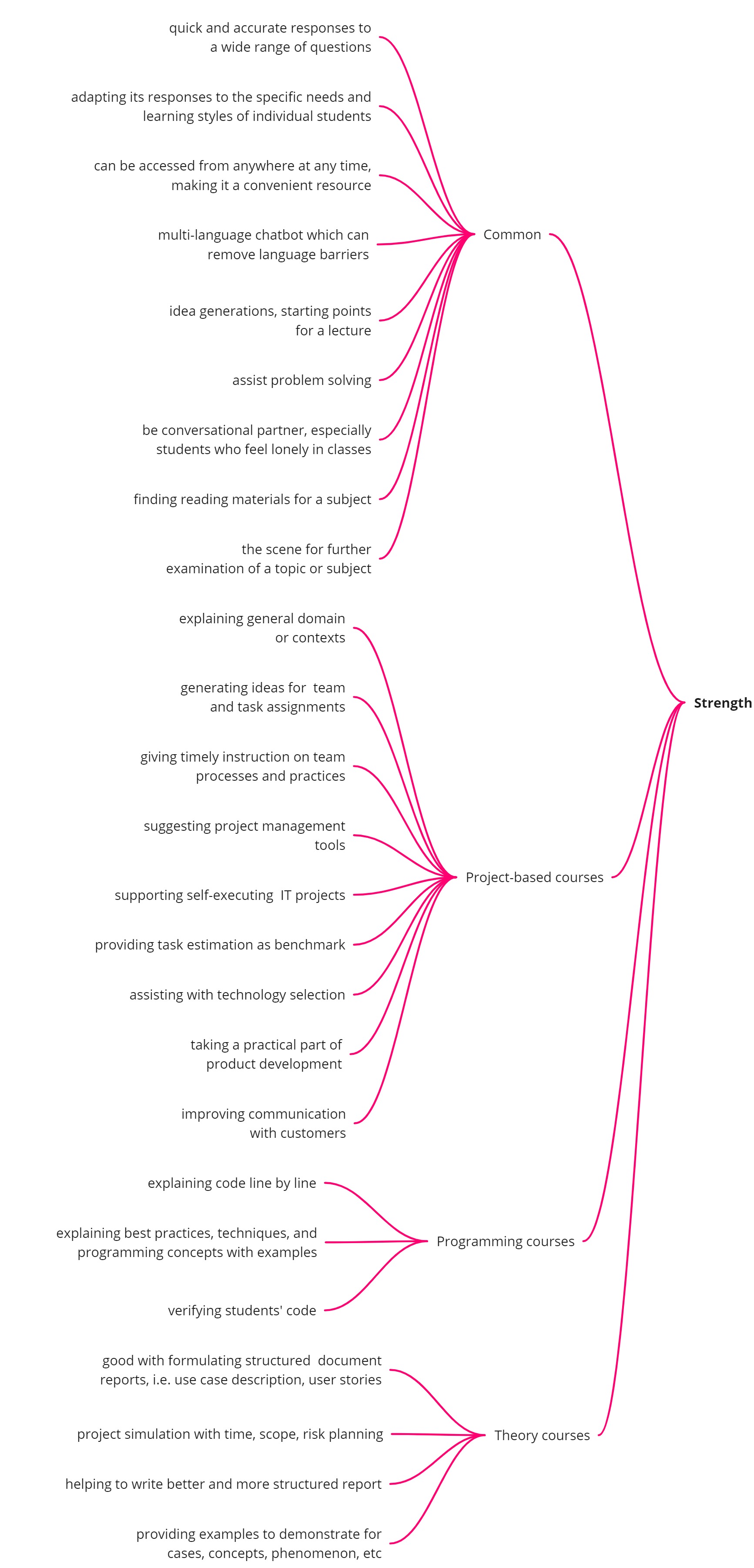}
    \caption{Strengths of ChatGPT in assisting IT education}
    \label{fig:enter-label}
\end{figure}

\subsection{Weaknesses} 
The weaknesses of ChatGPT in our education are described in four categories: common weaknesses, programming courses, project-based courses and theory courses. One common weakness of ChatGPT in education is that its responses are limited to the information provided in the question. This means that if a question is poorly formulated or lacks relevant details, the response from ChatGPT may not be satisfactory. Additionally, the model that ChatGPT is based on is trained on pre-existing data up to 2021, which means that it may not have up-to-date knowledge on certain topics. Answers from ChatGPT may also need to be verified by experts to ensure accuracy and prevent the spread of misinformation. Another weakness is the requirement for awareness and critical thinking of what the tool can provide, which students might not have. This could lead to false confidence about certain topics or subjects. There is also a repeatability issue, as different answers can be generated for the same question, which could lead to confusion for students. When data is limited, responses from ChatGPT can be shallow or incorrect. Finally, ChatGPT has limited domain knowledge, so questions that require a lot of context or background information may not receive meaningful answers without sufficient input from the user. Questions such as “how do I implement system X in organization Y?” would require feeding chatGPT with a lot of contexts in order to receive meaningful answers.

Regarding programming courses, tools like ChatGPT cannot provide hands-on experiences, such as coding challenges or real-world projects. Firstly, while it can provide information and guidance on programming concepts, students cannot develop their practical skills solely by using ChatGPT. For example, for a student who is learning about Machine Learning models, and needs to build the model, ChatGPT cannot provide it. Secondly, with lectures with a lot of visual representations, ChatGPT relies on natural language processing and may struggle to interpret visual information. For example, in a lecture about 3D modeling, ChatGPT may not be able to explain a particular aspect of the modeling process that is best understood through visual demonstration. Thirdly, unlike teachers, ChatGPT cannot tell industry-specific programming practices. While ChatGPT can provide information on general programming concepts, it may not be able to provide insight into specific practices used in a particular industry. For example, a student studying game development may need to learn about specific optimization techniques used in the gaming industry, which ChatGPT may not be familiar with.

Regarding project-based learning, where project contexts can change frequently, it is necessary to update the context information for the tool to catch up with project progress. For example, if a project in a software engineering class changes its requirements, students may need to make changes to their code. ChatGPT may not be able to keep up with these changes and may not be able to provide relevant advice to students. This is also because real-world settings can not be fully described as input for the tool. ChatGPT is based on pre-existing data and may not be able to provide practical advice for unique situations that occur in real-world settings. For example, a business student may need advice on how to handle a specific client situation that is not covered by the pre-existing data used to train ChatGPT.

\begin{figure}
    \centering
    \includegraphics[width=1\linewidth]{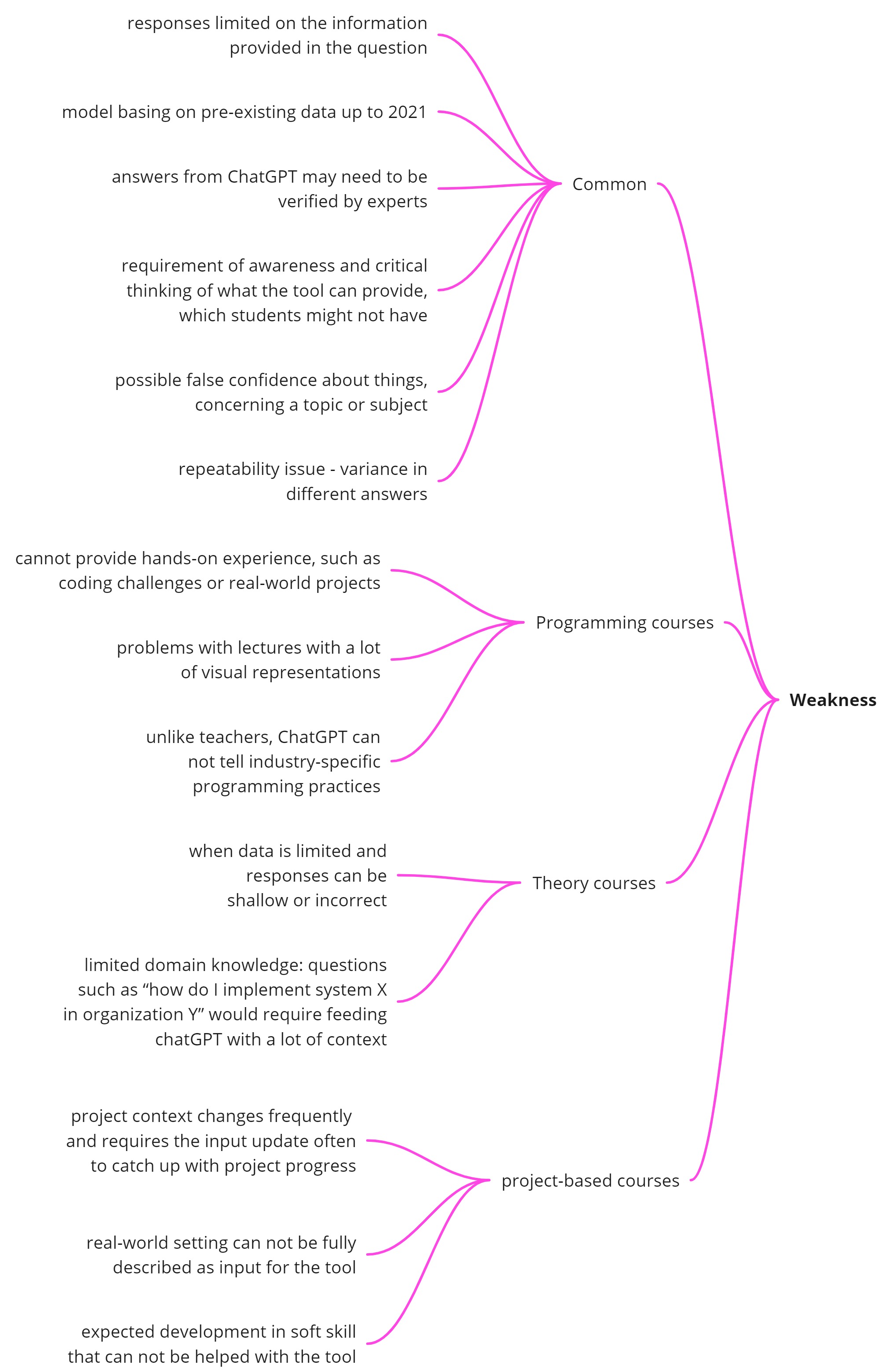}
    \caption{Weaknesses of ChatGPT in assisting IT education}
    \label{fig:enter-label}
\end{figure}

\subsection{Opportunities} 
The potential use cases for ChatGPT in education are diverse and beneficial for teachers and students. In theory and programming classes, ChatGPT can assist in structuring thoughts and learning activities. Students can utilize the chatbot to organize their ideas, identify knowledge gaps, and create mind maps or outlines of course reading materials.

Teachers can rely on ChatGPT to design curricula based on existing materials and receive suggestions for new content and topics. The chatbot can also aid in planning lecture schedules to ensure comprehensive coverage of necessary subjects. Additionally, ChatGPT can be employed as an educational tool to teach students about the advantages of using AI tools in their learning process. Students can actively engage with the chatbot to support their own learning endeavors.

Furthermore, ChatGPT is capable of organizing and reorganizing curriculum components based on personal requests, assisting students in structuring assignments, and providing outlines and content suggestions for reports. It can also educate students with little or no IT background by offering simple explanations and examples in a user-friendly manner. The chatbot can provide personalized advice on the best courses of study tailored to individual skills and experiences.

By combining ChatGPT with an active learning or flipped classroom approach, the tool can adapt to student's needs and facilitate a personalized and responsive learning experience. It can generate new assignments for the class by incorporating contextual data such as current events or student interests. Moreover, students can work at their own pace and receive feedback on their progress, making learning more flexible. Lastly, ChatGPT can effortlessly generate customized examples and code snippets to enhance students' understanding of CSS or JavaScript concepts.

In a programming course, the opportunities for assistance with source code include code explanation, bug fix, learning about concepts and code generation. ChatGPT can also be used as a personal assistant for students. It can help students manage their time, schedule, and assignments. Code explanation is one of the opportunities that ChatGPT can provide in education. ChatGPT can help students understand code snippets or programming concepts that they may have difficulty with. For instance, if a student is having difficulty understanding a specific code syntax or function, they can use ChatGPT to get an explanation of that code snippet or concept. ChatGPT can provide step-by-step instructions, examples, and explanations to help students understand programming concepts.
Bug-fixing is another opportunity that ChatGPT can provide in education. ChatGPT can help students debug code or find errors in their programming assignments. For instance, if a student has an issue with their code and is unable to identify the source of the problem, they can use ChatGPT to get help in debugging their code. ChatGPT can analyze the code, identify the problem, and suggest a solution to the student. 

\begin{figure}
    \centering
    \includegraphics[width=1\linewidth]{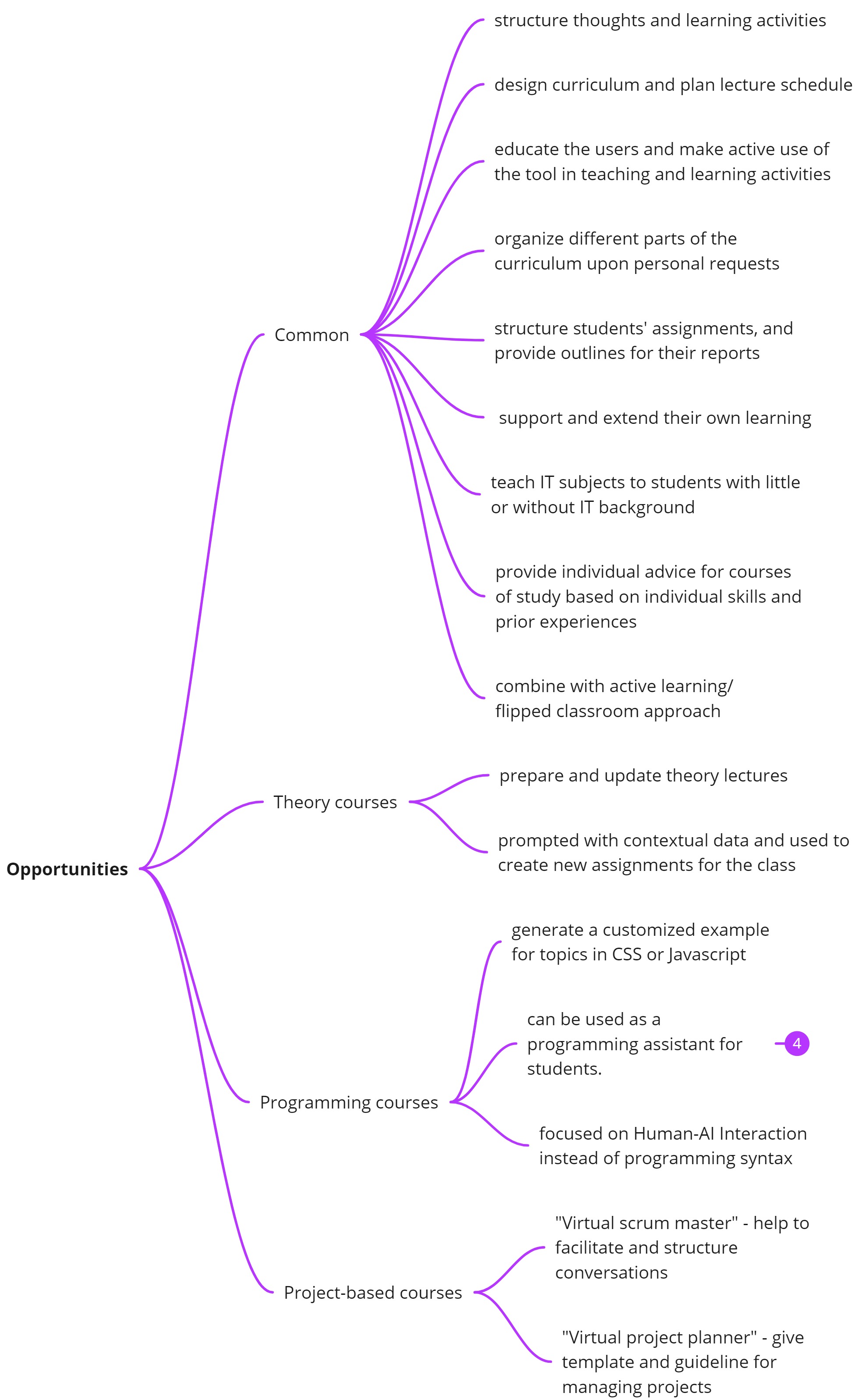}
    \caption{Opportunities of ChatGPT in assisting IT education }
    \label{fig:enter-label}
\end{figure}

\subsection{Threats} 
To realize the potential of chatbots, there are potential threats to consider. Figure 6 describes common threats and threats for programming courses and theory courses. 
Student may lose their ability to develop critical thinking. By relying too heavily on ChatGPT to generate ideas or provide answers, students may lose their ability to think independently and critically evaluate information. This can happen when students use the chatbot to access vast amounts of data without the required accountability, which can undermine academic integrity and personal growth. One common threat is the potential for cheating with the tool, as students may use ChatGPT to access a vast amount of data and information without the required accountability. Additionally, potential technical issues, such as server downtime, connectivity problems, or software bugs, could disrupt the learning process. Another potential threat is insufficient training with the tool, which may lead to misuse, such as students relying on code generation without understanding it. 
 
\begin{figure}
    \centering
    \includegraphics[width=1\linewidth]{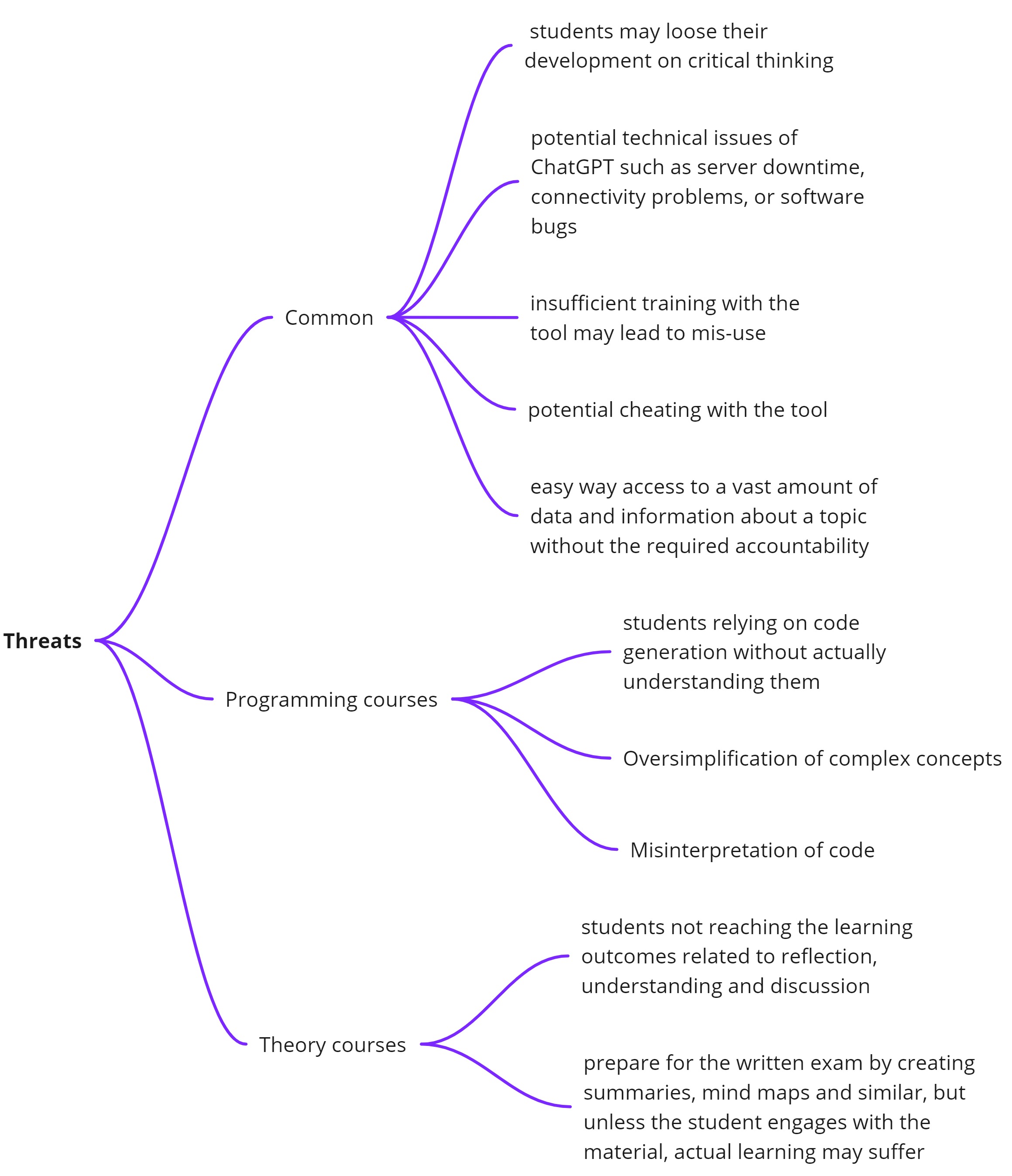}
    \caption{Threats of ChatGPT in assisting IT education}
    \label{fig:enter-label}
\end{figure}

In programming courses, this threat is especially significant, as students may misuse ChatGPT to generate code without understanding it. For example, in a computer science course, students may use ChatGPT to generate code for a programming assignment without understanding the code, resulting in a lack of mastery of programming concepts. Similarly, in a literature review task, students may use ChatGPT to generate summaries without engaging with the text, resulting in a superficial understanding of the material. Moreover, there is a risk of oversimplification of complex concepts, as ChatGPT may not always be able to provide in-depth explanations of complex programming concepts, resulting in a shallow understanding of the material. Misinterpretation of code is also a potential threat. 

In theory courses, ChatGPT can help prepare students for written exams by creating summaries and mind maps. Hence, it uncritically would lead to a shallower understanding and students not reaching the learning outcomes related to reflection, understanding and discussion. ChatGPT can help prepare for the written exam by creating summaries, mind maps and similar, but unless the student engages with the material, actual learning may suffer.
\section{Discussions and Conclusion} 
This section presents our implications to learners and educators, the possible changes to higher education and our future work. 
\subsection{Implications to learners and educators} 
With time and the advancement of technology, it is assumed that the usage of ChatGPT will be increased on both leaners’ and educators’ sides.  This will incur some changes in the education settings if it has not already disrupted the course \cite{qadir_engineering_2023,daun_how_2023,jalil_chatgpt_2023}. First, the teacher needs to be aware of the impacts, effects, and usage of ChatGPT in his/her course.  S/he needs to assess the scopes and possibilities, and strengths, weaknesses, opportunities, and threats of using ChatGPT in different teaching, learning, and assessing activities for his/her course.  For adopting successfully, the strengths and opportunities of using ChatGPT, learning activities and assessment methods might need to be revised and changed. Besides, threats are particularly important to address as they may jeopardize the overall learning outcome and the credibility of assessment in the course. Assessment methods and process needs to be revised and changed as well as exam question patterns and content and the focus of the test might need reassessed and changed in order to ensure that they hold values and integrity. Courses may also need to include instructions for students about how to use ChatGPT, where to use and where not to use, including the correct way of referencing ChatGPT work to avoid cheating and dishonesty. 
\subsection{Possible disruptive change in higher education} 
ChatGPT can improve education, as presented in the benefits in the result section \cite{qadir_engineering_2023,daun_how_2023,jalil_chatgpt_2023}. ChatGTP is used almost every day by students and several teachers are incorporating it into their teaching. ChatGPT has appeared with many benefits and challenges, just as search engines like Google did back in 1998. The launch of ChatGPT in November 2022 did not allow teachers to prepare for this technology. This prompt appearance of ChatGPT and fast-growing popularity among students has forced teachers to research, use and be up to date with the potential impact that this technology might have on their courses and teachings. As shown in nine IT courses at USN, several teachers have already implemented ChatGPT in their teaching, while other teachers have encouraged students to use it for learning and assignments. Teachers of these courses have expressed that the use of ChatGPT is unavoidable. As such, it is better to adapt to this new technology and incorporate it in different ways during teaching. Nevertheless, ChatGPT has led teachers to wonder how they can trust their students. 
Incorporating ChatGPT into a sustainable model has become a goal for several teachers. ChatGTP is re-inventing the teaching process, it’s transforming the way teachers deliver assignments and assessments. How to avoid plagiarism and cheating in projects and home exams has become an ongoing discussion among the teachers at USN. Since ChatGPT is unavoidable, it is paramount to create a participatory culture where students and teachers work together to examine the outcomes and implementation of this disruptive technology. In addition, the university should establish guidelines on the use of ChatGPT to ensure transparency, trustworthiness, and create a safe teaching environment.
\subsection{Future work} 
One potential avenue of exploration is to investigate the ethical implications of using ChatGPT as a teaching tool. As previously discussed, ChatGPT has several ethical issues, such as discrimination, no attribution, weak and arrogant character, and consent and privacy concerns. Future research work could examine how these ethical issues could impact the use of ChatGPT in IT education and develop strategies to address them. Additionally, the research could investigate the effectiveness of ChatGPT as a teaching tool in IT education, exploring its ability to engage students, provide personalized learning experiences, and enhance their understanding of complex topics.
Another area of future research work in adopting ChatGPT in IT education is exploring its potential impact on the job market. With the increasing demand for IT skills, there is a need for effective and innovative ways of teaching and training the future workforce. ChatGPT could offer a new approach to IT education that is more accessible, personalized, and engaging. Future research could explore how the adoption of ChatGPT in IT education could impact the job market, such as whether it could lead to a shift in the skills and knowledge required by employers or whether it could lead to an increase in demand for IT-related jobs. Additionally, the research could examine the potential benefits and drawbacks of ChatGPT in IT education, such as its impact on the quality of education, the role of teachers, and the cost-effectiveness of implementing this technology.

\bibliographystyle{splncs04}
\bibliography{main}

\end{document}